\documentclass[aps,twocolumn,pre,floatfix]{revtex4}
\usepackage{graphicx}\voffset=2cm
\begin {document}

\title{Fractal and Statistical Properties of Large Compact Polymers: a
Computational Study}

\author{Rhonald Lua*, Alexander L.  Borovinskiy*\dag, Alexander Yu.  Grosberg
*\ddag}

\affiliation{*Department of Physics, University of Minnesota,
Minneapolis, MN 55455, USA \\ \dag Present address:  University of California at San Francisco \\
\ddag Institute of Biochemical Physics, Russian Academy of
Sciences, Moscow 117977, Russia}
\date{\today}

\begin{abstract}
We propose a novel combinatorial algorithm for efficient
generation of Hamiltonian walks and cycles on a cubic lattice,
modeling the conformations of lattice toy proteins. Through
extensive tests on small lattices (allowing complete enumeration
of Hamiltonian paths), we establish that the new algorithm,
although not perfect, is a significant improvement over the
earlier approach by Ramakrishnan \emph{et. al.} \cite{Conf-des},
as it generates the sample of conformations with dramatically
reduced statistical bias.  Using this method, we examine the
fractal properties of typical compact conformations.  In
accordance with Flory theorem celebrated in polymer physics, chain
pieces are found to follow Gaussian statistics on the scale
smaller than the globule size.  Cross-over to this Gaussian regime
is found to happen at the scales which are numerically somewhat
larger than previously believed.  We further used Alexander and
Vassiliev degrees 2 and 3 topological invariants to identify the
trivial knots among the Hamiltonian loops.  We found that the
probability of being knotted increases with loop length much
faster than it was previously thought, and that chain pieces are
consistently more compact than Gaussian if the global loop
topology is that of a trivial knot.

\end{abstract}
\maketitle

\section{Introduction}

The dominant mood among the protein folding experts these days
seems to suggest that we are rapidly approaching the day when
experiments and theory - or, rather, simulations - will be ready
for direct quantitative comparison.  New generation experiments,
including single molecule ones
\cite{Fersht_Nature_2003,One_at_a_time,Lipman} provide the long
awaited insights into the folding paths.  New proteins are
discovered or invented exclusively with the goal to see their
folding on the time scale more accessible to simulations. In the
complementary drive, modern computer simulations
\cite{Shimada2,Shimada,Thirumalai}, particularly those employing
so-called distributed computing \cite{distributed}, not only
consider explicitly all atoms (although no explicit water), but
also rapidly improve in terms of the ways to treat forces involved
\cite{forces,Thirumalai2,Thirumalai3,Mayo}.  The impressive
episode of a theoretical prediction \cite{prediction} verified by
the experiment \cite{confirmation} is celebrated \cite{PandeNV} as
the sign of approaching new level of integration between theory
and experiments.

In our opinion, all these shining achievements only highlight once
again how badly we need a better insight into the simple
fundamentals of folding. Just as the decoding of genomes does not
cancel, but strengthens the pressing need of orders of magnitude
higher throughput reading systems, in the same way deeper
understanding of the underlying simple physical principles behind
protein folding remains one of the most needed pieces of the
puzzle.  With this point in mind, in this work we try to address
deeper the properties of the simplest caricature proteins, namely,
lattice ones.

Of course, in our work with simple toy models we should keep an
eye on the progress of more elaborate studies.  What do they teach
us?  In the opinion of the present authors, what stands out as a
common lesson in all computational studies of protein folding is
the central importance of the interplay between two trivial facts
- the first is that proteins are polymers, and the second is that
they are compact (globular) polymers.  Very highly non-trivial
geometry comes with these facts
\cite{singular_derivatives,vas_knots,dimension,foam,Kussell1,Kussell2}.
This opinion was also explicitly formulated in the recent News and
Views \cite{PandeNV}.

What do we know about compact polymer conformations?  Protein data
bank contains large and rapidly growing collection of
conformations.  Should there be any general principle behind these
conformations? Many authors are looking for such principles,
either biological (selection-driven), or physical, geometrical,
etc.  Not even starting to discuss the existing theories, their
advantages and disadvantages, we would like to point out that such
discussion remains premature as long as properties of
\emph{random} compact conformations are not understood well.
Indeed, having no insight into the majority of arbitrary
conformations, we cannot judge how non-random are the
conformations in protein data bank.  For instance, there are
relatively few knots in native proteins
\cite{Mansfield1,Mansfield2,knotted_protein,Nature_knots_protein};
is it because unknotted conformations are somehow biologically
selected, or are they physically preferable for, e.g., folding -
or alternatively, maybe, what seems to be "few" for us is, in
fact, statistically expected number of knots in compact
conformations of the given length?  Currently, we cannot answer
this.

The theory of random compact conformations is well developed on
the mean field level (see, e.g., in the book \cite{RedBook}). This
is the theory of homopolymer globules, because they are
entropically dominated by the most typical conformations.  Major
conclusion of the mean field theory is that chain segments inside
the globule follow Gaussian statistics, and do not exhibit any
signs of order.  This conclusion is in sharp contradiction with
the statements in the literature \cite{Dill,Banavar1,Banavar2}
that compactness of the conformation may favor elements of
secondary structures, such as $\alpha$-helices and $\beta$-pins.

Computationally, the problem of compact conformations is closely
related to that of Hamiltonian walks on the graphs.  We remind the
reader that the concept of a Hamiltonian walk was introduced by
Hamilton in connection with famous Euler problem of K\"onigsberg
bridges: the task was to find the Sunday promenade passing every
one of the seven bridges, never returning to the already visited
place.  In general, Hamiltonian walk on an arbitrary graph can be
defined as a walk which visits every site on the graph once and
only once. If our graph is, say, $\ell \times m \times n$ piece of
the cubic lattice in $3D$, then Hamiltonian walk on such graph is
the same as maximally compact conformation of the polymer filling
$\ell \times m \times n$ domain.

Enumeration of Hamiltonian walks on graphs is well known problem
in combinatorics. Of course, the best possible statistics is
achieved by exhaustive enumeration of all Hamiltonian walks.  This
is possible for rather short polymer chains only: for the chains
with $27$ monomers filling $3 \times 3 \times 3$ of the cubic
lattice \cite{Shakhnovich}, and also for $36$- and $48$-mers,
filling $3 \times 3 \times 4$ and $3 \times 4 \times 4$ segments,
respectively \cite{enum_latt334}. Obviously, these chains are far
too short to address statistics and fractal structure of the
typical conformation.

Short of exhaustive enumeration, other methods to generate larger
compact conformations have been suggested.  The most
straightforward Monte Carlo chain growth methods \cite{Monte
Carlo} are totally inefficient for long compact chains, because of
catastrophic explosion of rejected looped conformations. Transfer
matrix approach put forward by
\cite{Jernigan1,Jernigan2,Jernigan3} is very efficient for the
chains filling an elongated domain $\ell \times m \times n$, where
one of the dimensions, say $n$, may be arbitrarily large.
Unfortunately, to remain within computational tractability, two
other dimensions, $\ell$ and $m$, must be small, not greater than
$2$ or $3$.  An alternative approach, suggested in
\cite{Conf-des}, is free of this limitation.  It employs
combinatorial techniques of two-matching and patching of bipartite
graphs.  Unfortunately, we found that this method generates
conformations in a heavily biased way.

The objective of our work is three-fold. First, we report the
improvements to the algorithm by  Ramakrishnan \textit{et al}
\cite{Conf-des}.  We must mention at once that even the improved
method is not free of biases; however, it is significantly better
in this respect than the original approach \cite{Conf-des}.
Second, we investigate the properties of the generated compact
conformations (Hamiltonian walks) and cycles against the polymer
length. The largest walks generated have the size $22\times
22\times 22$.  Third, we examine the topology of maximally compact
closed loops, including the loop length dependence of the trivial
knot probability, as well as the local fractal structure of the
typical conformation for both averaged loop and the loop which is
trivial as a knot.

The article is organized as follows. The proposed new algorithm is
formulated in details in the next section \ref{sec:methods}.  The
results of the implementation of this algorithm are presented in
section \ref{sec:implement}.  The topological properties of the
compact knots are considered in the section \ref{sec:knots}.  At
the end, we discuss the conclusions from our study in section
\ref{sec:conclusion}.

\section{Methods}\label{sec:methods}

\subsection{Construction of the lattice graph}

We performed our simulations on $L \times L \times L$ cubic
lattices with $L=2,3, \ldots , 22$, but our algorithm applies for
any finite regular bipartite graph.  The graph is called
\emph{bipartite} if two colors suffice to paint it in such a way
that every two neighboring vertexes have different colors. Chess
board is a good example of a bipartite graph; three vertices
connected as a triangle is an example of a graph which is not
bipartite.   We call the graph, or lattice, \emph{even} or
\emph{odd} if the total number of vertexes, $N$, and, therefore,
the length of Hamiltonian walk, is even or odd, respectively.
Obviously, $L \times L \times L$ cubic lattice is the bipartite
graph, with $N=L^3$; it is even or odd for even or odd $L$,
respectively.

The following very simple theorem can be established regarding the
Hamiltonian walks on bipartite graphs.  If a bipartite graph is
colored, say, using black and white colors, then the walks on this
graph necessarily step from black to white or vice versa.
Therefore, every Hamiltonian walk on an even lattice starts and
ends on different colors, while on the odd lattice its ends occupy
the vertices of the same color.  Moreover, on the odd lattice one
of the colors can be called \emph{major}, because there are more
sites of one color than the other ($(N+1)/2$ vs. $(N-1)/2$).  We
shall call this simple statement {\bf the chess board theorem}.   One of
the conclusions of the chess board theorem is that the Hamiltonian
cycles are impossible on the odd lattices, because every cycle on
the bipartite graph must contain equal number of sites of both
colors.

\begin{figure}[ht]
\centerline{\scalebox{0.85} {\includegraphics{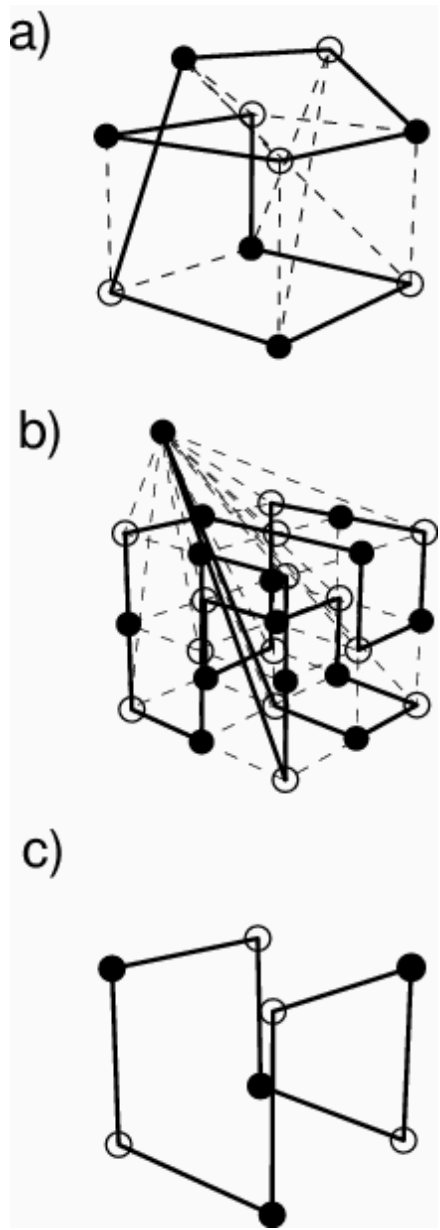}}}
\caption{The construction of the lattice graphs for generation of
a) Hamiltonian walk on even lattice; b) on odd lattice; c)
Hamiltonian cycle. The walks are drawn as solid lines and the
edges of the lattice graphs as dash lines.} \label{fig:outoflatt}
\end{figure}

From the discussion above, it may seem that generation of
Hamiltonian walks on odd and even lattices, and generation of
Hamiltonian cycles on even lattices, are three very different
problems which should be treated separately.  In fact, they can
all be reduced to one another by the trick proposed in the article
\cite{Conf-des}.  Let us introduce \emph{extended} graph by adding
some \emph{out-of-lattice} vertices using the following rules:
\begin{itemize}  \item In case of even lattice, we add two out-of-lattice
vertices of different colors (see Fig. \ref{fig:outoflatt}a).  We
connect them to each other, and each of them - to all the lattice
vertices of the opposite color.  \item In case of odd lattice, we
add only one out-of-lattice vertex, which is colored minor color
and connected to all major color "real" vertices (Figure
\ref{fig:outoflatt}b).
\end{itemize}
Constructed this
way, extended lattices are always even.  Therefore, all we have to
do is to generate Hamiltonian cycles on the even lattices.  As
soon as that problem is addressed, we can generate Hamiltonian
cycle on the extended lattice and obtain open Hamiltonian walk by
just removing the out-of-lattice vertices.

\subsection{The algorithm}

The original combinatorial algorithm by Ramakrishnan et al
\cite{Conf-des} consists of two steps.  First, it generates some
configuration of sub-cycles and sub-chains  with dead ends on the
lattice by means of \emph{two-matching} procedure; second, it
transforms these pieces into a single Hamiltonian walk using
another procedure called \emph{patching}.  The main novelty of our
algorithm is that the formation of sub-cycles and sub-chains is
forbidden, and we always generate the single Hamiltonian cycle on
the extended lattice graph.  Thus, patching stage becomes
unnecessary.  We explain in the Appendix \ref{sec:bias}, why the
formation of small loops and sub-chains in the original method
\cite{Conf-des} biases sampling of the Hamiltonian walks.

\begin{figure}[ht]
\centerline{\scalebox{0.75} {\includegraphics{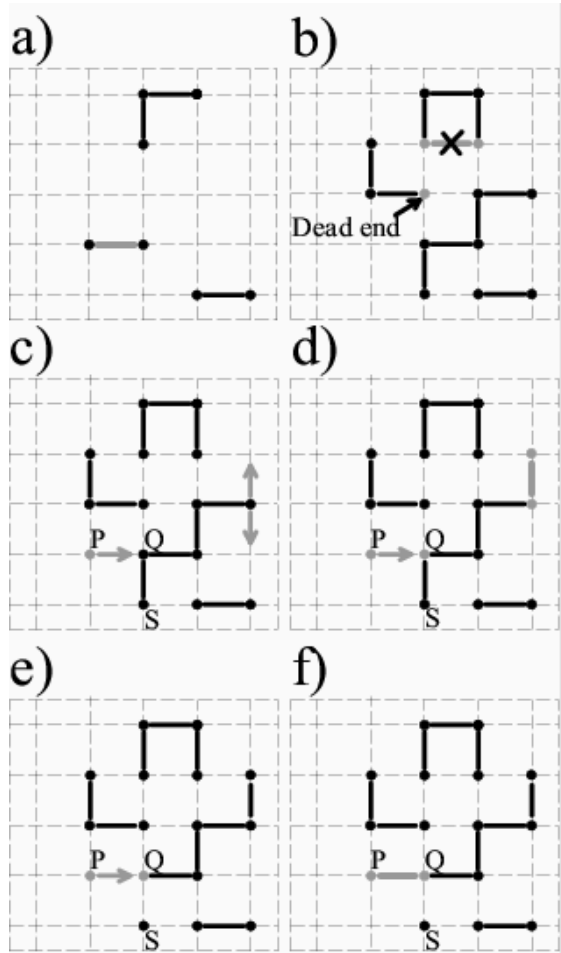}}}
\caption{Schematic representation of the application of the
algorithm.  For simplicity, steps of the algorithm are shown in
two dimensions. See text for further explanations.}
\label{fig:algorithm}
\end{figure}

The algorithm works by placing links on the lattice graph.  At the
beginning, the lattice graph contains no links.  Then, algorithm
starts placing links randomly, connecting randomly chosen
neighboring vertices (Figure \ref{fig:algorithm}a).  Every time a
new link is chosen, we check whether it forms an unwanted small
subtitle or a dead end (Figure \ref{fig:algorithm}b), and the link
is rejected if this happens. (The only little exclusion from the
general rule is required for an even lattice, where the first link
is always drawn between the out-of-lattice vertices, and this link
is never removed on the later steps of the algorithm.)   The
algorithm stops when all vertices of the graph are
\emph{saturated} by two links each, and the links form a
Hamiltonian cycle.  The obvious difficulty is that randomly chosen
vertex frequently cannot be linked to its randomly chosen
neighbors, because the latter is already saturated (Figure
\ref{fig:algorithm}c). This is the situation in which
\emph{two-matching} is applied.

Two-matching starts from picking up a vertex, $P$, which is
currently either not connected, or has only one incoming link.
Then, its random neighbor $Q$ is chosen as an opposite end of the
new link.  If $Q$ belongs to some linear sub-chain, we peak up
randomly one of the links incoming to it and follow this direction
along the sub-chain.  When the sub-chain terminus is found, it is
investigated for the possibility to be connected with one of its
neighbors.  For each vertex, all the non-saturated neighbors
ending the sub-chain are placed on the special list.  The
neighbors are not included in the list if linking with them leads
to the formation of sub-cycles or dead ends (Figure
\ref{fig:algorithm}d). Then, a random vertex from the list (of
course, if the list is not empty) is chosen, and the new link is
drawn (Figure \ref{fig:algorithm}e). The growth of the sub-chain
is followed by the switching of the links incident on $Q$. The
link such as $QS$ (see Figure \ref{fig:algorithm}f; the link
opposite to the one pointing to the end just elongated) is removed
and the new link $PQ$ is drawn, subject to the following two
conditions: i) the vertex $P$ is still unsaturated after the
elongation of the sub-chain; ii) linking the vertices $P$ and $Q$
does not produce subtitle or dead end.  Depending on the success
of two processes contributing to the two-matching, the number of
links on the graph increases by one, remains the same, or
decreases by one.  In our simulations, the latter case was rare
and did not slow the process too much.

The new links are placed on the graph until finally a single cycle
passing once and only once through every vertex of the graph
(including the out-of-lattice ones) is formed.
\begin{figure}[ht]
\centerline{\scalebox{0.75} {\includegraphics{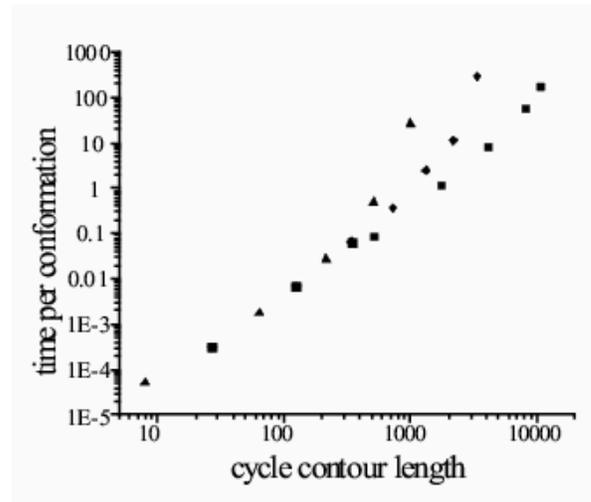}}}
\caption{Performance of the algorithm for generation of
Hamiltonian walks and cycles on cubic lattices. The results for
even walks are shown as triangles, for odd walks as diamonds, for
cycles as squares.} \label{fig:performance}
\end{figure}

\subsection{Algorithm performance test}

We implemented the algorithm described just above to generate
linear polymer chains up to the size $12 \times 12 \times 12$ on
even lattices, up to $15 \times 15 \times 15$ on odd lattices, and
the compact cycles of the sizes up to $22\times 22 \times 22$. On
the lattices larger than mentioned this algorithm becomes
exponentially slow, however, for the investigated lattices, we
found the CPU time necessary to generate one chain conformation
demonstrates power law dependence on the length of the walk, $N$.
The effectiveness of our algorithm executed on the Pentium III 1.1
GHz PC is demonstrated in Figure \ref{fig:performance}. The run
time scales approximately as $N^{2.1}$ for both linear polymers
and cycles for the moderate chain lengths. This is slower than
performance reported in \cite{Conf-des} for the original algorithm
($\sim N^{1.1}$).  This is the price we must pay to ensure fair
sampling. Still, our algorithm allows to generate compact polymer
chains within the length range of several orders of magnitude.

\subsection{Topological aspects}

There exists abundant literature on computational studies of the
knot composition of non-compact closed chains, starting with the
pioneering work of Vologodskii et al
\cite{Vologodskii,VologodskiiReview,Michels,Koniaris,Shima,DeguchiTsuru}.
These studies are mostly motivated by the intent to model closed
circular DNA.  There are much fewer studies made with compact
chains \cite{Tesi,Mansfield}, although the question of knots in
proteins is widely considered a puzzle
\cite{Mansfield1,Mansfield2,knotted_protein,Nature_knots_protein}.

We should particularly emphasize the work by Mansfield
\cite{Mansfield}, where he addressed knots in Hamiltonian cycles
on the cubic lattice.  What we add here to his analysis is we pull
it to significantly longer loops, which turns out to be essential,
and we also study the statistics of the sub-chains in the loop
whose overall global topology is fixed.

As in all previous works, we applied the theory of knot invariants
to determine the knot-type of a given conformation. Knot
invariants are mathematical objects that serve as a 'signature' of
the knot-type. As a signature, knot invariants are, unfortunately,
not unique to a given knot. The use of the appropriate types and
number of knot invariants yields only a good likelihood that the
knot has been identified correctly.  This likelihood is high, in
certain cases unity, if the number of crossings in the knot
projection could be reduced to a sufficiently small number.  The
difficulty we have to face here is that compact conformations have
typically very large numbers of crossings on the projection.

In this work, we calculated for a knot $K$ three invariants - the
Alexander polynomial ($\Delta (t)_{K}$) evaluated at a certain
value of $t$, ($\Delta (-1)_{K}$), the Vassiliev invariant of
degree two ($v_{2}(K)$), and the Vassiliev invariant of degree
three ($v_{3}(K)$) - as was also done in \cite{DeguchiTsuru}. A
connection is made between a conformation and its knot-type if the
invariants calculated from the projection of the conformation
coincide with the invariants associated with the knot-type.

\begin{figure}[h]
\centerline{\scalebox{0.5}{\includegraphics{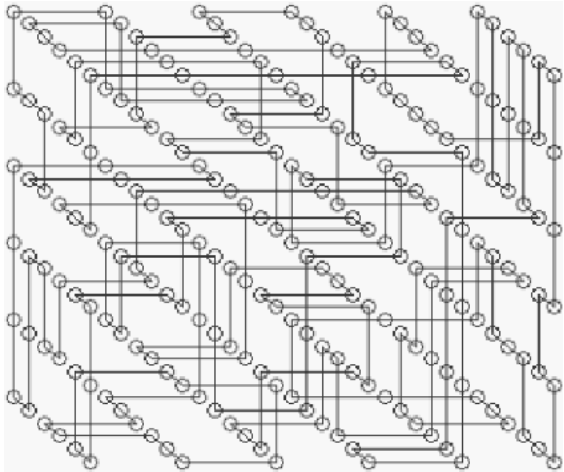}}}

\caption{Projected nodes and links of a $6 \times 6 \times 6$
conformation. The knot formed is a
trefoil.\label{cap:Projected-nodes-and}}
\end{figure}

In order to illustrate the necessity of topological invariants in
identifying even the simplest knots, including the trivial knot
(which is an unknot) we show Figure \ref{cap:Projected-nodes-and}.
In fact, the loop shown in this figure is a trefoil knot, but it
is virtually impossible to realize this fact by eye.

Thus, after a compact conformation has been generated, the
procedure for determining its knot-type involves the following
steps: (1) Generate Plane Projection; (2) Preprocess Projection;
(3) Compute Knot Invariants from Projection; (4) Match
Conformation with Knot-type using Table \ref{tab:knots}.

\subsubsection{Preprocessing Projection}

The goal of preprocessing the projection is to simplify the knot
by reducing the number of intersections or crossings of the
projected links. The intuitive local 'moves' that can accomplish
this simplification are called Reidemeister moves (see, for
instance, \cite{Mura}).  Given the very complicated nature of
typical compact conformations, we resort to combinations of
Reidemeister moves, compounded, or 'macro', as discussed in
\cite{Kauffman}.

For large conformations, a further simplification can be achieved
by first 'inflating' the conformation before taking the
projection. A less dense conformation leads to a significant
reduction of crossings. In fact, this was done for $14\times
14\times 14$ conformations before the Vassiliev invariants were
evaluated.

\subsubsection{Computing Knot Invariants}

An algorithm for computing the Alexander polynomial $\Delta
(t)_{K}$ is presented clearly in \cite{Vologodskii} and will not
be discussed any further here. Suffice it to say that the
algorithm requires the construction of an 'Alexander' matrix from
the knot projection, with dimension equal to the number of
crossings. The determinant is subsequently calculated after
setting $t$ to $-1$ to obtain the single number $\Delta (-1)_{K}$.

The geometrical origin of this invariant may be traced to
'linking' numbers calculated from a set of closed curves. These
closed curves are associated with a 'Seifert surface' whose
boundary is the knot \cite{Mura}.

The calculations for the Vassiliev invariants
($v_{2}(K)$,$v_{3}(K)$) are presented as diagrammatic formulas in
\cite{Polyak}. These formulas operate on a Gauss diagram, or
equivalently on a Gauss code for a knot $K$. The set of Vassiliev
invariants may be considered as a generalization of the Gauss
integral formula for the linking number.

As mentioned earlier, it is possible for two distinct knots to
have the same set of knot invariants. However, we expect that the
false identification of a knot would be rare. For instance, the
set of three knot invariants for the trivial knot is distinct from
those of (prime) knots with $10$ minimum crossings or fewer ($249$
knots in all) in their projection.

\begin{table}\caption{Values of knot invariants for a few knots.}
\begin{center}\begin{tabular}{|c|c|c|c|c|}
\hline KNOT& Alexander, & Vassiliev, & Vassiliev,&
CHIRAL?\\ & $\left|\Delta (-1)_{K}\right|$ &$v_{2}(K)$ &$\left|v_{3}(K)\right|$& \\
\hline \hline $0_{1}$& 1& 0& 0& NO\\ (Trivial) &&&& \\ \hline
$3_{1}$& 3& 1& 1&
YES\\
\hline $4_{1}$& 5& -1& 0&
NO\\
\hline $5_{1}$& 5& 3& 5&
YES\\
\hline $5_{2}$& 7& 2& 3&
YES\\
\hline
\end{tabular}\label{tab:knots}\end{center}
\end{table}

\section{Results: compact chains}\label{sec:implement}

\subsection{Statistics for the small lattices}

As a first test of our algorithm, we compare the statistics of
generated random samples with the results of exhaustive
enumeration for $2\times 2\times 2$ and $3\times 3\times 3$ cubic
lattices.

\begin{figure}[ht]
\centerline{\scalebox{0.30} {\includegraphics{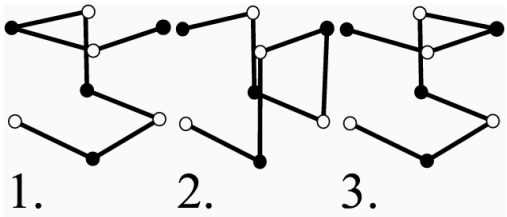}}}
\caption{There are three symmetrically unrelated conformations
possible on $2\times 2 \times 2$ cubic lattice.}
\label{fig:conformations}
\end{figure}

For the $2\times 2\times 2$ lattice the task is easy, because
the complete list  consists of only 3 symmetrically unrelated
Hamiltonian walks. These walks are shown in the Figure
\ref{fig:conformations}. The unbiased algorithm should generate
each  of these 3 conformations with  probabilities $1/3$. We
generated  samples of $100000$ walks using our algorithm and
using the original algorithm of Ramakrishnan et al
\cite{Conf-des}.  The average fractions of different walks in
generated samples obtained with both algorithms are shown in Table
\ref{222}. Clearly, the algorithm \cite{Conf-des} fails this test;
the reasons of its failure are explained in the Appendix
\ref{sec:bias}.

\begin{table}
 \caption{The average fractions of different $2\times
2\times 2$ conformations in generated samples obtained with two
algorithms.}
\begin{tabular}{|c|c|c|c|}\hline
&\multicolumn{3}{|c|}{Conformation} \\ \cline{2-4} Algorithm & 1 &
2 & 3\\ \hline Ramakrishnan et al \protect\cite{Conf-des}& 0.278 & 0.358
& 0.364 \\
\hline
present & 0.328 & 0.328 & 0.344\\
\hline
\end{tabular}\label{222}
\end{table}

For the $3\times 3 \times 3$ lattice a little more elaborate
procedure is necessary.  Suppose, there are some $M$ conformations
(for instance, $M = 103346$ for $3 \times 3 \times 3$ lattice
\cite{Shakhnovich}), and suppose we repeatedly apply one and the
same algorithm to generate a number $K$ of Hamiltonian walks.
Apart from glitches with the random number generators, subsequent
applications of the algorithm are statistically independent.
Therefore, for every conformation $i$ there is the occurrence
probability $p_i$.  For the unbiased algorithm, $p_i = 1/M$; in
general, $\epsilon_i = p_i - 1/M$ measures the bias.  To examine
this bias, we compute the distribution $m_k$ - for every number of
appearances $k$, $m_k$ is the number of conformations that
appeared $k$ times in $K$ trials.  Obviously, $m_k$ is normalized
such that $\sum_{k=0}^K m_k = M$.  Since appearances of every
particular conformation are binomially distributed, we have
\begin{equation} m_k = \sum_{i=1}^M p_i^k
(1-p_i)^{K-k} \frac{K!}{k! (K-k)!} \ , \label{eq:multipeak}
\end{equation}
where the summation runs over all conformations.   From here, it
is not difficult to find that, first of all, the average (over all
conformations) appearance number is $\overline{k} = K / M$, it is
independent of a bias.  The information about the bias is
contained in further moments of the distribution.  Specifically,
we consider the further cumulants of the distribution of
$\epsilon_i$: variance
\begin{eqnarray} \langle \epsilon^2 \rangle_{cum} & \equiv &  \langle
\epsilon^2 \rangle = \nonumber \\
& = & \frac{1}{K^2} \left[\overline{ \left( k - \overline{k}
\right)^2} - \frac{K}{M} \right] \ , \label{eq:bias}
\end{eqnarray}
skewness
\begin{eqnarray} \langle \epsilon^3 \rangle_{cum} & \equiv &  \langle
\epsilon^3 \rangle = \nonumber \\
& = & \frac{1}{K^3} \left[ \overline{\left( k - \overline{k}
\right)^3} -3 \overline{ \left( k - \overline{k} \right)^2} + 2
\frac{K}{M} \right] \ , \end{eqnarray}
and kurtosis
\begin{eqnarray} \langle \epsilon^4 \rangle_{cum} & \equiv &  \langle
\epsilon^4 \rangle - 3 \langle \epsilon ^2 \rangle^2 = \nonumber \\
& = & \frac{1}{K^4} \left[ \overline{ \left( k - \overline{k}
\right)^4} - 6 \overline{ \left( k - \overline{k} \right)^3}  +
\right. \nonumber \\ && \left.+ 11 \overline{ \left( k -
\overline{k} \right)^2} - 3 \overline{ \left( k - \overline{k}
\right)^2}^2 - 6 \frac{K}{M} \right] \ ,
\end{eqnarray}
where averaged (over all conformations) powers of $\epsilon$ are
defined according to
\begin{equation} \langle \epsilon^n \rangle = \frac{1}{M} \sum_{i=1}^M
\epsilon_i^n \end{equation}

\begin{figure}[ht]
\centerline{\scalebox{0.75} {\includegraphics{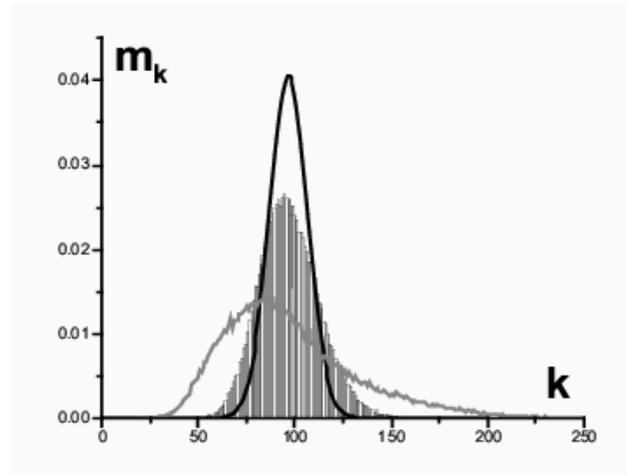}}}
\caption{The computed distributions of not symmetrically related
conformations on $3\times 3 \times 3$ lattice by the frequency of
generation obtained by our method (columns) and method of article
\cite{Conf-des} (grey line) compared with the distribution
expected for the unbiased algorithm (black line).  Here, $k$ is
the number of times a conformation appeared in $K=10000000$
trials, while $m_k$, for every $k$, is the number of conformations
which appeared $k$ times.  The number of different Hamiltonian
walks on $3\times 3 \times 3$ lattice is $M=103346$.}
\label{fig:ergodicity}
\end{figure}

We generated two samples of $K= 10000000$ Hamiltonian walks by
means of our algorithm and the one of the article \cite{Conf-des}
and compared the appearance of different Hamiltonian walks in
these samples. The obtained distributions $m_k$ for both
algorithms are shown in Fig. \ref{fig:ergodicity}. The
distribution (\ref{eq:multipeak}) for the unbiased $\epsilon =0$
case (when it is simply a Gaussian with the mean $K/M$ and
variance also $K/M$) is also presented in the same Figure. The
parameters of the computed distributions are summarized in the
Table \ref{333}.

\begin{table}
\caption{The parameters of computed distributions of conformations
on $3\times 3\times 3$ lattice obtained with two algorithms.}
\begin{tabular}{|c|c|c|c|}\hline &&& \\
Algorithm & $\frac{\left( \langle \epsilon^2 \rangle_{cum}
\right)^{1/2}}{1/M} $ & $\frac{\left( \langle \epsilon^3
\rangle_{cum} \right)^{1/3}}{1/M} $ & $\frac{\left( \langle
\epsilon^4 \rangle_{cum} \right)^{1/4}}{1/M} $ \\ &&& \\ \hline
Ramakrishnan &
&  
&  
\\ et al \protect\cite{Conf-des} &$0.34$&$0.34$&$ 0.35$ \\
\hline
present & $0.12$ 
& $ 0.09$ 
& $0.21$ 
\\
\hline
\end{tabular}\label{333}
\end{table}

As the data indicate, our algorithm produces the distribution,
which is close to the expected unbiased result. The distribution
shape is very closely Gaussian, which means the bias is weak. At
the same time, the algorithm of the article \cite{Conf-des} showed
poor results and produced the distribution, which is essentially
skewed. This demonstrates strong biases of that method.

A not so good news about our algorithm
is that the width of the distribution is still larger than expected for unbiased sampling.
Given the width of the distribution we can estimate the bias
from formula (\ref{eq:bias}), $\epsilon=1.2\times
10^{-6}$.  This signals certain bias, about $10 \%$, in the
generation of Hamiltonian walks.   However, the bias is
small, and certainly much smaller than for the
previous algorithm. In what follows, we shall
examine the statistics of Hamiltonian walks generated by our
algorithm and neglecting its bias.


\subsection{Statistics of segments and loops in generated
walks}\label{sec:testing_Flory_theorem}

By the statistics of segments we understand the following. Imagine
a long polymer compressed in a very compact state, and suppose a
part of the chain, some $\ell$ monomers long, is labeled.  For
instance, it may be deuterated.  Then, we can study the
conformation of the labeled segment.  Is it collapsed, with the
overall size scaling as $\ell^{1/3}$?  Is it extended, with
end-to-end distance scaling as $\ell^{1}$?  Does it exhibit any
signs of regularity, such as helical structure of some sort?  Or
is it purely random, yielding Gaussian statistics with the size
scaling as $\ell^{1/2}$?  This is the question we want to address
here.

To begin with, let us remind the major conclusions of the mean
field theory (see, e.g., review in the book \cite{RedBook}). This
theory suggests that labeled chain segment behaves similarly to
the labeled chain in a macroscopic polymer melt or concentrated
solution of different chains.  Therefore, it obeys Flory theorem
\cite{Flory,Edwards,DeGennesBook}.  To appreciate the highly
non-trivial statement of the Flory theorem, one has to realize
first of all that either labeled chain in the concentrated melt,
or labeled $\ell$-segment in the globule, is subject to the volume
exclusion constraint: trivially, other monomers cannot penetrate
the volume occupied by any given monomer.  As it is well known in
polymer physics, volume exclusion leads to polymer swelling, with
significant correlations between monomers, and with chain size
scaling $\ell^{\nu}$, $\nu \approx 0.588 \approx 3/5$.  It is not
difficult to realize that the presence of surrounding chains in
the melt, or surrounding parts of the same chain in the globule,
leads to some effective attraction between labeled monomers. Flory
theorem says that this attraction exactly compensates the excluded
volume effect.  In other words, surrounding polymer medium shields
excluded volume effect, leaving labeled chain with Gaussian
statistics and the size proportional to $\ell^{1/2}$.  This
screening is sometimes called Edwards screening, it is similar to
Debye screening in plasma.

What is the range of $\ell$ in which Gaussian scaling $\ell^{1/2}$
is expected?  Of course, $\ell$ must be larger than the effective
Kuhn segment - which is equal to unity for the lattice model.
Another restriction, relevant for the globule and not for the
melt, is that labeled segment as a whole should be away from
globule boundaries, or surfaces.  Assuming globule size about
$N^{1/3}$ for the globule of density one and the chain of $N$
monomers, we arrive at the condition $\ell^{1/2} < N^{1/3}$, or $1
< \ell < N^{2/3}$.

Although this is not very important for the present study, we
would like to digress to inform the reader that even within the
mean field level, there are delicate corrections to the simple
picture as described above.  To understand this, one should think
of an auxiliary problem of a Gaussian polymer without excluded
volume confined in a cavity with impermeable walls.  Under such
conditions, chain adopts a conformation with density peaked at the
middle of the cavity and with density almost vanishing at the
cavity walls \cite{RedBook}.  The contrast between this
theoretical model and the real globule with flat distributed
internal density suggests that self-consistent field acting inside
the globule not only compresses the chain, acting like a cavity,
but also pulls the monomers from globule center to the periphery.
This pull slightly perturbs Gaussian statistics of the sub-chains,
particularly those located nearby the globule boundary.
Computationally, we shall not look into this delicate effect in
our present study.

Thus, we compute the mean square end-to-end distances of the
segments of Hamiltonian walks:
\begin{equation}
\langle R^2 (\ell)\rangle = \frac{1}{K(N-\ell)} \sum_j^K \sum_i^{N-\ell}
\left( {\vec r}_{i+\ell}^{(j)} - {\vec r}_i^{(j)} \right)^2 \ ,
\label{eq:subchain_size}
\end{equation}
where $\ell$ is the contour length of the segment of the walk (in
units of steps), $K$ is the total number of walks in the
sample, $N$ is the length of the walk, ${\vec r}_i^{(j)}$ is the
position vector of the vertex visited $i$-th in the $j$-th walk.

\begin{figure}[ht]
\centerline{\scalebox{0.50} {\includegraphics{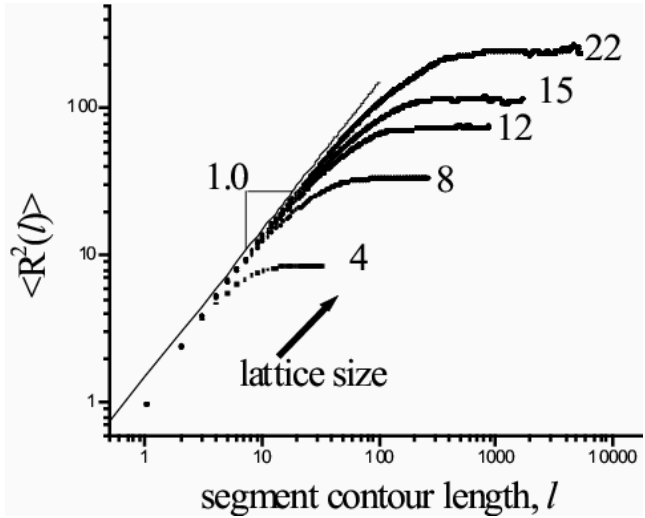}}}
\caption{Mean square end-to-end distance of the segments of
Hamiltonian walks vs. the lengths of segments is shown for the
lattices of  different sizes. The curves for linear walks and
cycles on the lattices $4\times 4\times 4$, $8\times 8\times 8$
and $12\times 12\times 12$ coincide.  } \label{fig:segments}
\end{figure}

The results for the samples of Hamiltonian walks of different
lengths are presented in Figure \ref{fig:segments}.  In good
agreement with mean field theory, on the scales smaller than
$N^{2/3}$ the walks obey Flory theorem \cite{Flory} and the
average distance between the segment ends scales such that
$\langle R^2 ( \ell ) \rangle \sim \ell$. We would like to note here
that Flory theorem does not tell us anything about the prefactor of this
scaling. Fitting on the statistics of the lattice polymer cycles of the
size $22\times 22\times 22$ suggests the prefactor to be equal $\approx
1.5>1$. For the polymer chain
without excluded volume it is exactly equal to $1$. Therefore, the
excluded volume effectively increases the Kuhn  segment length.

On the scales $\ell \sim
N$, the walk starts feeling the confinement by the lattice
borders, and $\langle R^2 (\ell )\rangle$ levels off.

\begin{figure}[ht]
\centerline{\scalebox{0.75} {\includegraphics{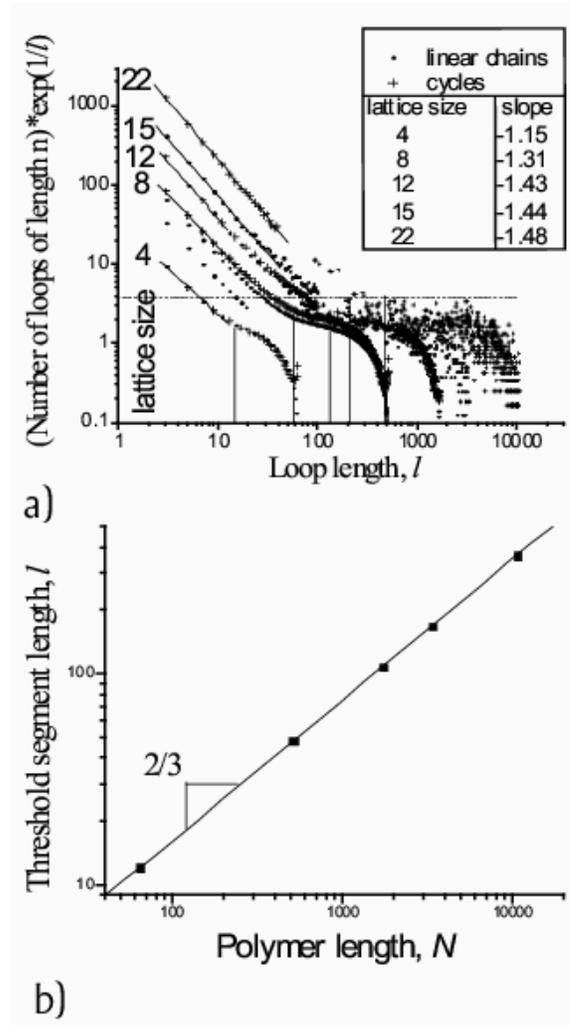}}}
\caption{(a)The average number of loops of various contour length
in generated Hamiltonian walks on the lattices of different size.
Vertical lines display the cross-over values of $\ell$ at which
looping probability saturates. Horizontal dash line corresponds to
the predicted saturation level for the $22\times 22\times 22$
walks. (b) The dependence of the cross-over value of $\ell$ on the
polymer length.} \label{fig:loops}
\end{figure}

Another measure of the agreement between statistics of Hamiltonian walks
and Flory theorem is the looping probability.
The Figure \ref{fig:loops}a shows how often the loops of different
contour lengths appear in the Hamiltonian walks.  Here, we say
that the walk makes a loop of the length $\ell$, if after visiting
site with the coordinates ${\vec r}_i$ it visits one of this site
neighbors in exactly $\ell$ steps.  What does the mean field
theory have to say about these loops?

As we saw for the statistics of end-to-end distances, on the
scales $\ell < N^{2/3}$, the Hamiltonian walks are Gaussian. Then,
the probability distribution of their end-to-end vectors ${\vec
R}$ must obey Gaussian law $\sim \ell^{-3/2} \exp \left[ - {\vec
R}^2 / \ell \right]$.  For the loop, $R=1$. Therefore, average
number of loops of the contour length $\ell$ should decay as
$\ell^{-3/2}\exp(-1/\ell)$ with growing $\ell$. That is why the
number of loops on the vertical axis of the Figure
\ref{fig:loops}a is weighted by the factor of $\exp(-1/\ell)$. We
can express surprise that power law $\ell^{-3/2}$ comes so slowly
and  appears only at large $N$ (see the table on the inset to Fig.
\ref{fig:loops}a).

We can also check cross-over value of $\ell$ and how it depends on
$N$. Vertical lines on the Figure \ref{fig:loops}a mark the
characteristic segment lengths at which the cross-over takes place
for the polymer chains of different length. And Figure
\ref{fig:loops}b shows the dependency of these threshold values on
the polymer length $N$. It is clearly seen that $\ell$ scales as
$N^{2/3}$.

On the larger scales, $\ell > N^{2/3}$, the probability to find
the loop of length $\ell$ saturates and becomes practically
independent of $\ell$.  To estimate its constant value, we can
resort to the following argument.  The random  walk of a length
greater than $N^{2/3}$ hits the borders of the lattice.  The
end of the longer walk may be found in any lattice site with
nearly equal probability $1/N$.  Since the loop formation
condition is met by $\langle z \rangle$ of sites neighboring to
the loop starting site, the loop probability is about $\sim \left.
\langle z \rangle \right/ N$.   Here, $\langle z \rangle$ is the
mean coordination number of the lattice (which takes into
account that the sites on the surface have fewer neighbors than
those in the bulk). At the same time, there are $N - N^{2/3}
\approx N$ such loops possible, therefore, there must be about
$\langle z\rangle$ loops of each length found in every walk.
Indeed, the horizontal dash line on the Figure \ref{fig:loops}a
corresponding to  $\langle z\rangle$ of the compact walk of the size
$22\times 22\times 22$ reasonably estimates the number of long loops in
the globule of this size.

The results presented in Figure \ref{fig:loops} are in full
agreement with the theory, both in terms of the power law decay
($\ell^{-3/2}$) at moderate $\ell$, the range of the cross-over
($\ell \sim N^{2/3}$), and the constant levels at large $\ell$
($\langle z\rangle$).

\subsection{Correlation between ends in Hamiltonian walks}

It is an interesting question in the theory of polymer globules,
whether the ends of the polymer chain are effectively independent
of each other in terms of their positions inside globules, or they
repel (attract) due to the conditions of the connectedness and
compactness of the chain. If the end of the chain is located in
the bulk of the globule, there may be entropic cost associated
with the rearrangement of the parts of the chain surrounding it
due to necessity to keep the compactness of the globule. This
local rearrangement of the polymer chain may affect the
probability of the other end  to locate in the vicinity.
Effectively, this may lead either to the attraction, or to the
repulsion of the ends of the chain.  Theoretically, this issue
remains currently unclear \cite{KardarOrland}.

To check on the existence of such effective interaction between
chain ends, we calculate the end-end correlation coefficient for
the samples of generated Hamiltonian walks.  This quantity is
defined via the formula
\begin{equation}
c = \frac{ \langle x_{1} x_{2} \rangle }{\sqrt{ \langle
x_{1}^2\rangle \langle x_{2}^2\rangle } },
\end{equation}
where $x_1$ and $x_2$ are the $x$-coordinates of the two chain
ends, $\langle \ldots \rangle$ means averaging over all sampled
walks.  For simplicity, we place coordinate system origin in the
center of the cube, such that $\langle x_1 \rangle = \langle x_2
\rangle = 0$.  Due to the symmetry, correlations coefficients for
$y$ and $z$ coordinates are the same as for $x$, while all the non
diagonal elements (such as $\langle x_1 y_2 \rangle$ etc.) vanish.

\begin{figure}[ht]
\centerline{\scalebox{0.75} {\includegraphics{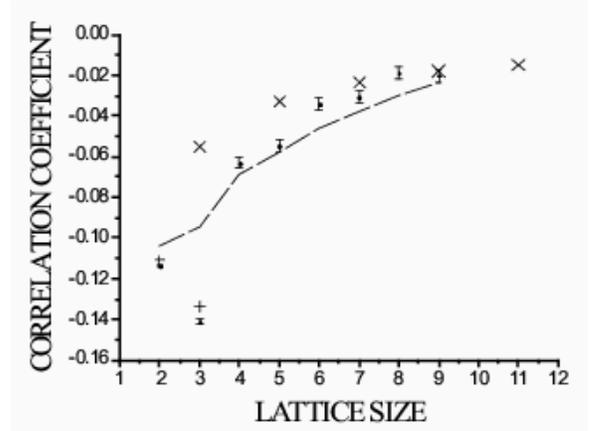}}}
\caption{Mean diagonal end-end correlation coefficients for the
Hamiltonian walks on the lattices of different sizes. The data of
exact calculation for $2\times 2 \times 2$ and $3\times 3 \times
3$ lattices are shown as $+$. The data of exact calculation of
correlation coefficients for the random pairs of dots on the odd
lattices obeying the excluded volume condition and the chess board
theorem are shown for the comparison as $\times$. The results
obtained from generation of the walks with algorithm of the work
\protect\cite{Conf-des} are shown as the dashed line.}
\label{fig:corr}
\end{figure}

The results obtained from the simulations on the lattices of the
size $L=2,3, \ldots, 10$ are presented in Figure \ref{fig:corr}
along with the data of the exhaustive enumeration  for the
$2\times 2\times 2$ and $3\times 3\times 3$ lattices and the exact
results for the disconnected ends model (which, due to the chess
board theorem, is only meaningful for odd lattices; for even
lattices, two ends must be on the oppositely colored sites, and,
therefore, are not correlated at all). The results for the small
lattices are very close to exact (whereas the original algorithm
\cite{Conf-des} produces significant systematic errors). This is
another good suggestion that our algorithm has weaker bias than
that of the work \cite{Conf-des}.

The fact that correlation coefficient is negative indicates that
there is some effective repulsion between the chain ends. This
effect decreases and supposedly goes to zero with increasing of
lattice size. Moreover, correlation between ends very rapidly
approaches correlation between disconnected points subject only to
excluded volume condition. This observation suggests that even the
small repulsive correlation between chain ends is mostly due to
the benign excluded volume effect of the terminal monomers, and
chain connectivity provides only faint, although also repulsive,
contribution (probably mostly due to excluded volume of monomers
next to the terminal ones).

\section{Results: compact loops and their knots}\label{sec:knots}

\subsection{Average Crossing Number}

\begin{figure}[h]
\centerline{\scalebox{0.35}{\includegraphics{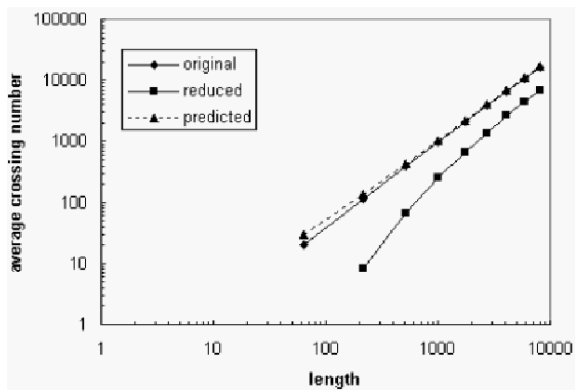}}}

\caption{Average crossing numbers in the knot projection, before
and after preprocessing with Reidemeister moves, together with
mathematical prediction. These were plotted against the size
(length $N=L^{3}$) of the conformation, from $L=4$ to
$L=20$.\label{cap:Average-crossing-numbers}}
\end{figure}

Figure \ref{cap:Average-crossing-numbers} displays the average
number of crossings in the plane projection of a conformation,
together with the reduced number and mathematical prediction, for
the range of sizes $L=4$ to $L=20$. The crossing numbers are
plotted against the length (number of monomers) $N=L^{3}$.

The prediction
\begin{equation} C=\left(\frac{L^{3}}{3(L-1)^{2}}-1\right)\frac{L^{3}}{3}
\end{equation}
for the average crossing number of an $L\times L\times L$
conformation follows from the assumption that every segment upon
projection in some 'vertical' direction produces crossings with
all segments above and below it inside the cylinder of the
cross-section unity. In this sense, the result for the average
crossing number is trivial. However, it is interesting to note
that for large $L$, the expression for the average crossing number
scales as $C=L^{4}=N^{\frac{4}{3}}$, which is reminiscent of a
'four-thirds power law' relating crossing number and 'rope length'
for tight knots \cite{Buck96,Canta,BuckNature}.  This suggests
that this four-thirds power law does not reflect on any intimate
properties of tight knots, except their overall space filling
character.

From the average crossing number, one could get an idea of how the
amount of computational resources involved in the calculation of a
knot invariant, say Alexander, scales with conformation size. The
Alexander invariant entails computation of the determinant of a
$C\times C$ matrix. Naively using Gaussian elimination,
computation time would roughly scale as $C^{3}=N^{4}$.

\subsection{Knot Probabilities}

\begin{figure}[h]
\centerline{\scalebox{0.35} {\includegraphics{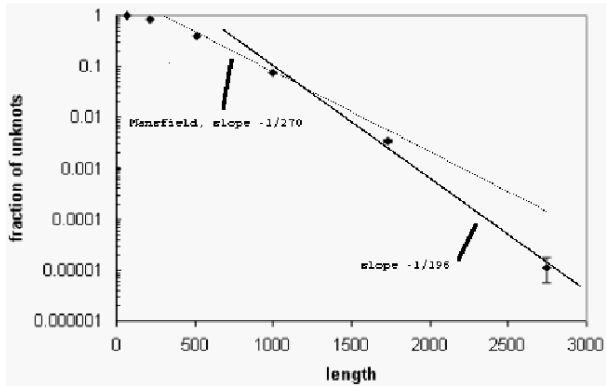}}}

\caption{Trivial knot probabilities for conformations of size
$L=4$ to $L=14$. The thinner dotted line represents Mansfield's
\protect\cite{Mansfield} fit to his data points.
\label{cap:Unknotting-probabilities-for}}
\end{figure}

Figure \ref{cap:Unknotting-probabilities-for} displays our results
for the fraction of conformations (of a given size $N=L^{3}$)
which are unknotted. For each $L$ from $4$ to $12$, $10^{5}$
conformations were generated. The last data point for the largest
conformation we were able to analyze ($14\times 14\times 14$)
represents $4$ trivial knots out of $350000$ conformations.

Since the total number of conformations of the length $N$ grows
exponentially with $N$, it is not a surprise that the probability
of a trivial knot decays exponentially with $N$
\cite{theorem1,theorem2}.  Accordingly, computational data on
trivial knot probability are customary fit to exponential.  In our
case, the exponential fit to the (last three) data points yielded
an estimate for the unknotting probability as a function of $N$,
$\sim \exp (-N /196)$, as shown in Figure
\ref{cap:Unknotting-probabilities-for}.

Previously, there were some works measuring knotting probabilities
for lattice polygons in confined geometries \cite{Tesi,Mansfield}.
In particular, Mansfield \cite{Mansfield} has examined knots of
compact Hamiltonian cycles on a lattice - the same problem we
consider here. However, these authors use one invariant, the
Alexander polynomial, in their computations (although Mansfield
\cite{Mansfield} evaluated Alexander polynomial at $10$ different
values of $t$). This is understandable, as the Vassiliev
invariants are a relatively recent discovery \cite{Polyak}, in
particular the invention of explicit and computationally
implementable formulas for their evaluation. Moreover, we were
able to analyze larger conformations: the work \cite{Mansfield}
examined $N \leq 1000$, while we consider $N$ up to $14^3=2744$,
almost three times larger.

Mansfield's fit to his results ($\exp (- N / 270)$) is shown in
the thinner, dotted line in Figure
\ref{cap:Unknotting-probabilities-for}.  Importantly, our results
for $N \leq 1000$ agree well with both the results and the fit by
Mansfield \cite{Mansfield}.  However, examination of larger $N$
leads us to revise the estimation of characteristic length $N_0$
in $\exp (-N/N_0)$ from $N_0 \approx 270$ to $N_0 \approx 196$.
Moreover, our result for $N_0$ may turn out an overestimate, and
real $N_0$ may eventually be found even smaller than $200$.
Indeed, the leading source of inaccuracy in our results is due to
the incomplete set of topological invariants.  This can lead to
errors of assigning the trivial knot status to some loops which
are in fact not trivial knots.  Such errors contaminate our
trivial knot sets with non-trivial knots, leading to the
\textit{overestimate} of trivial knot probability, and this effect
only increases with growing $N$, because at small $N$ it is much
less likely to meet a non-trivial knot confused with trivial one
by our set of knot invariants.  Thus, we conclude that the trivial
knot probability for compact polymers goes as
\begin{equation} w_{\rm compact, \ trivial} \simeq \exp \left( - N
/ N_0 \right) \ ,  \ \ \ \ \ N_0 \alt 196 \ . \end{equation}

This result is essential for several reasons.  We have shown in
the section \ref{sec:testing_Flory_theorem} that the sub-chains
inside the sufficiently big compact globule behave somewhat like
Gaussian polymers, with $R^2(\ell)$ proportional to $\ell$ despite
the obvious presence of volume exclusion constraint.  This fact,
consistent with Flory theorem, leads to the traditional
understanding that the chains in the melt as well as sub-chains in
the globule \textit{are} Gaussian.  From this, it would then be
logical to assume that the trivial knot probability for them
should also be the same as for corresponding Gaussian polymers,
and not the same as for the swollen self-avoiding polymers.  We
remind that the trivial knot probability for Gaussian polymers,
that is, for polygons of $N$ segments with no volume exclusion,
also follows the exponential law $\exp (- N / N_0)$, with $N_0$
varying from about $350$ for Gaussian random polygons (in which
all segments have Gaussian distributed lengths)
\cite{DeguchiTsuru} to about $260$ for regular polygons (made of
length $1$ segments) \cite{Koniaris,Nathan}.  For the
self-avoiding polymers, the value of $N_0$ is even larger
\cite{Shima,EXP(-28)}.  Our result now indicates that in regard to
the knot forming ability of the polymer, chain compaction not only
screens away the excluded volume, reducing $N_0$ from its value
for "thick" polymers to that for "thin" ones, but produces the
much more dramatic effect, decreasing $N_0$ significantly below
its Gaussian value.  In brief, compact polymers, although they
satisfy Flory theorem, are \textbf{not} Gaussian for topological
purposes, they are much (exponentially) more prone to forming
knots.

\begin{figure}[h]
\centerline{\scalebox{0.35} {\includegraphics{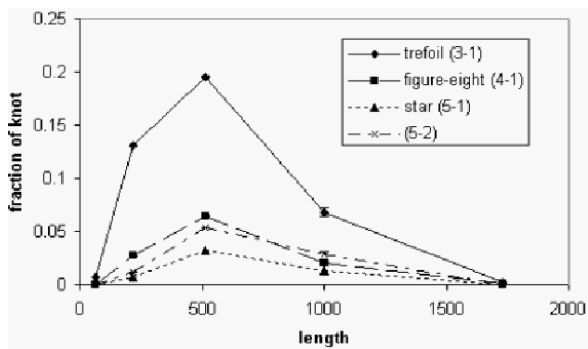}}}

\caption{Probabilities of occurrence of a few
knots.\label{cap:Probabilities-of-occurrence}}
\end{figure}

The Figure \ref{cap:Probabilities-of-occurrence} displays the
probabilities of some non-trivial knots in compact loops as the
function of the loop length.  Similar to the studies made with
non-compact chain models (see, e.g., \cite{DeguchiTsuru,Shima}),
the probability to obtain any particular knot starts from $0$ at
small $N$, then reaches a maximum at some finite value of $N$, and
then decreases and asymptotically approaches to $0$ with further
growth of $N$.  As in other cases, the qualitative explanation of
this tendency is clear.  When $N$ is small, the loop might be too
short to form a given knot.  In fact, for the lattice model, it is
clear that for every knot there is a finite value of $N$ below
which this knot cannot be formed at all, so its probability is
exactly $0$ (for instance, the shortest loop capable of forming a
non-trivial knot on the cubic lattice has $N=24$ segments).
However, even for significantly larger $N$ there might still be
relatively few conformations to realize the given knot, and that
yields low probability.  At the other end, when $N$ is exceedingly
large, there are great many knots which can be comfortably formed,
and their number keeps increasing with $N$, yielding a decaying
probability to locate the given knot.  We should emphasize that
the results presented in Figure
\ref{cap:Probabilities-of-occurrence}, although qualitatively
reasonable, have somewhat preliminary character, because our use
of the restricted set of topological invariants at the very high
crossing numbers may lead to inaccurate knot assignments.

\subsection{Statistics of segments and loops in trivial knots}

In this section, we want to address the following problem.
Consider a sub-chain of some length $\ell$ which is large, but
much smaller (in a proper sense) than the entire globule.  Suppose
further that the chain as a whole is closed, so it is a loop, and
that this loop is a trivial knot.  On the one hand, since $\ell
\ll N$, it seems that the sub-chain has no way to "know" what are
the global topological properties of the entire loop.  On the
other hand, it is also obvious that the property of being a
trivial knot is not a local but a global property of the loop. In
some loose sense, we can say that since the entire loop has no
knots, there is no way the sub-chain of the length $\ell \ll N$
may have knots.  Of course, to speak about knots in a sub-chain we
should somehow decide how to close its ends; what we are saying
here is that the sub-chain of an unknotted loop must not have
knots under the majority of natural ways to connect its ends. This
logic then seems to suggest that the sub-chain may tend to be
swollen compared to its random walk size $\ell^{1/2}$, based on
the analogy with loops in unrestricted space in which trivial
knots are known to swell \cite{Shima,Nathan,3_5_uzla}.  However
attractive, this logic at least does not exhaust the problem,
because if sub-chain sizes were to scale as $\ell^{\mu}$ with $\mu
\geq 1/2$, then these sub-chains would strongly overlap in the
overall compact globule, making it difficult to avoid making knots
between the sub-chains.  All these inconclusive arguments are
presented here in order to motivate the problem:  how does the
sub-chain size (say, end-to-end distance) scale with the sub-chain
length if the sub-chain is buried deeply inside a collapse trivial
knot?

\begin{figure}[h]
\centerline{\scalebox{0.40}{\includegraphics{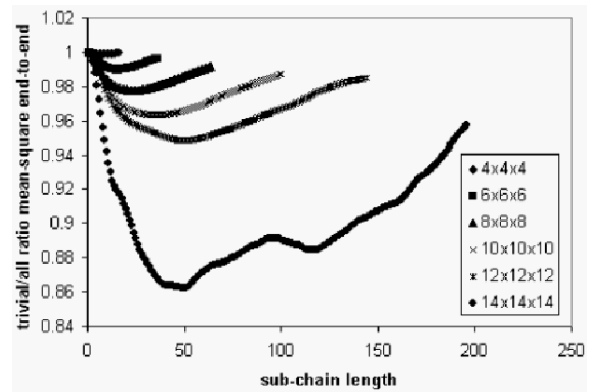}}}

\caption{Ratio of sub-chain mean-square end-to-end distance in
trivial knots and in all loops versus number of links in the
sub-chain. For the chain of the length $N=L^3$, filling $L\times
L\times L$ cube, results were plotted up to $L^{2}$.
\label{cap:Ratio-of-sub-chain}}
\end{figure}

\begin{figure}[h]
\centerline{\scalebox{0.40}{\includegraphics{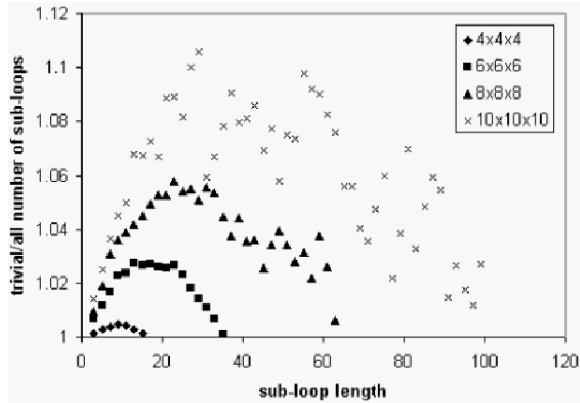}}}

\caption{Ratio of number of sub-loops in trivial knots and in all
loops, versus number of links in the sub-loop. Results for $L=12$
and $L=14$ were not plotted due to excessive 'noise'. This result
complements figure
\ref{cap:Ratio-of-sub-chain}.\label{cap:Ratio-of-subloops}}
\end{figure}

Measurements of mean-square end-to-end distance (defined similarly
to Eq. (\ref{eq:subchain_size})) were made on sub-chains (segments)
of compact chain conformations with trivial knots and on
sub-chains of all conformations regardless of knot-type. The
results (figure \ref{cap:Ratio-of-sub-chain}) show that sub-chains
of trivial knots are smaller or more compact compared to
sub-chains of all knots. A similar result was also obtained for
the gyration radius, which is another measure of size. (The
extrema in the plots are, of course, effects of the finite size of
the conformations.)

Measurements of the number of sub-loops formed in each
conformation were also made (Figure \ref{cap:Ratio-of-subloops}. A
loop is formed when monomers not connected by a link are next to
each other in space). The result for the number of sub-loops in
conformations with trivial knots, compared to the number of
sub-loops in conformations regardless of knot-type, is in complete
agreement with the previous results:  since sub-chains are more
compact in overall trivial knots, they are more likely to form
sub-loops.

These results should be contrasted to the corresponding results
for gyration radius of (entire) non-compact rings, which indicate
that trivial knots in such rings are, on average, larger compared
to all other knots \cite{Shima,Nathan}.  This is understood
\cite{3_5_uzla} based on the argument that there are very compact
conformations available for non-trivial knots are included in the
average over all loops and are excluded from the average over
trivial knots only.  Clearly, for this ensemble of unrestricted
loops, trivial knots remain swollen compared to the
all-loops-average not only on the level of entire polymer, but
also on the level of the sub-chains.  In fact, this effect is
expected to be scale-invariant at the length exceeding the
characteristic knotting length $N_0$ \cite{3_5_uzla}.  Based on
this comparison, we can conclude that it must be significantly
more difficult to confine a trivial knot loop into a small volume
than to realize a similar confinement of a phantom polymer, either
a chain or a loop.  Indeed, to compress a trivial knot one has to
reduce its entropy by forcing all the sub-chains to shrink.  This
means, confinement entropy for the trivial knot is a volume
effect, it scales as $N$ in thermodynamic limit.  It must be
compared with confinement entropy of usual polymers which only
scales as $N^{2/3}$ \cite{RedBook}.  This conclusion of the
increased stiffness of trivial knots compared to other loops is
consistent with the data of the work \cite{Nathan} on the
probability distributions of the unrestricted loop sizes: with
decreasing overall loop size, this probability decreases much
sharper for trivial knots than for averaged loops.

Although short of a proof, our results are consistent with the
hypothesis of a "crumpled globule," which was formulated many
years ago \cite{crumpled}, and which remains in the rank of
hypothesis till today.

\section{Conclusion}\label{sec:conclusion}

We formulated the new combinatorial algorithm for generation of
Hamiltonian walks and cycles on the cubic lattices. This algorithm
reduces biases compared to the previously known methods. The
presented algorithm performs well on generation of the large
compact self-avoiding walks.

We employed the proposed generation algorithm to verify Flory
theorem in its applicability to the random compact chains.  We
found that the statistics of the sub-chains inside the large
globule approaches Gaussian, as predicted by Flory theorem, for
sufficiently long polymers.   Unexpectedly, this happens at rather
large values of chain length $N$, about $10^5$.  Although it is
not entirely clear what is the most reasonable numerical
correspondence between $N$ for the lattice toy model and the
number of residues a the real protein, it is safe to question the
direct applicability of Gaussian statistics for the interior of
even large protein globules.  On the other hand, it should be
understood that the deviations from Gaussian statistics found for
modest $N$ compact chains are really small, and unless one is
interested in sophisticated scaling analysis, they provide very
reasonable qualitative fit to the data.

Using knot invariants, we were able to identify the trivial knots
and the first few knots in a sample of loop conformations.  We
found that the probability of trivial knot in a compact
conformation is significantly smaller than was previously
believed, and that it is much smaller than for the corresponding
Gaussian polymer.  This suggests that there should be an abundance
of knots in a random sample of compact conformations.  We have
also found that global restriction that the loop as a whole is a
trivial knot has a dramatic statistical effect on the
conformations of all sub-chains, making them significantly more
compact than for other loops.

Our results suggest that low propensity of knots in real proteins
might in fact be a statistically significant fact requiring an
explanation, although it seems too early to speculate what this
explanation might be, whether it is related to the physics of
folding, or to some functional properties of proteins, or to some
aspect of their evolution.

\acknowledgements

Authors acknowledge useful discussions with T.Deguchi and N.Moore.

Computations for the present work were performed using Minnesota
Supercomputing Institute facilities.

This work was supported in part by the MRSEC Program of the
National Science Foundation under Award Number DMR-0212302.

\appendix

\section{Is the new combinatorial algorithm
unbiased?}\label{sec:bias}

The building of the Hamiltonian walk on the lattice with the help
of some combinatorial algorithm can be viewed as the process of
labeling the edges of the lattice according to some rules (as
\emph{two matching, patching} or other procedures). One of the
rules is that none of the lattice nodes may have more then two
labeled edges incident on it. There are different configurations
of the labeled edges possible on the lattice. We now would like to
consider the space of all the possible such configurations. Such
space itself can be represented as a graph, in which every
configuration of labeled edges is a vertex, and two vertices are
connected if and only if the corresponding configurations differ
only by the labels of one lattice edge. Such space includes
configuration in which none of the edges is labeled. We call such
a configuration \emph{root}.
 The space can be divided into  the following subspaces:\\
i) configurations of labeled edges at which some of the lattice nodes do
not have incoming labeled edges (disconnected nodes);\\
ii) configurations containing multiple sub-cycles and sub-chains,
all the lattice nodes have two incident  labeled edges except the
ends of the sub-chains. No  new lattice edge can be labeled.
(Such configurations the algorithm \cite{Conf-des} used to start
\emph{patching} procedure);\\
iii) Hamiltonian cycles. \\
The configuration space is
schematically shown in the Figure 9. As an illustration we display
different configurations possible on the extended $2\times 2\times
2$ lattice.

An arbitrary combinatorial algorithm building a Hamiltonian walk starts
from the root node of the configuration space graph, then performs
random walk along some path on the graph, and finishes its work at some
node of subspace (iii). For  the algorithm to be unbiased, the number of
all possible paths leading to each node in the  subspace (iii) should be
equal.

\begin{figure}[ht]
\centerline{\scalebox{0.75} {\includegraphics{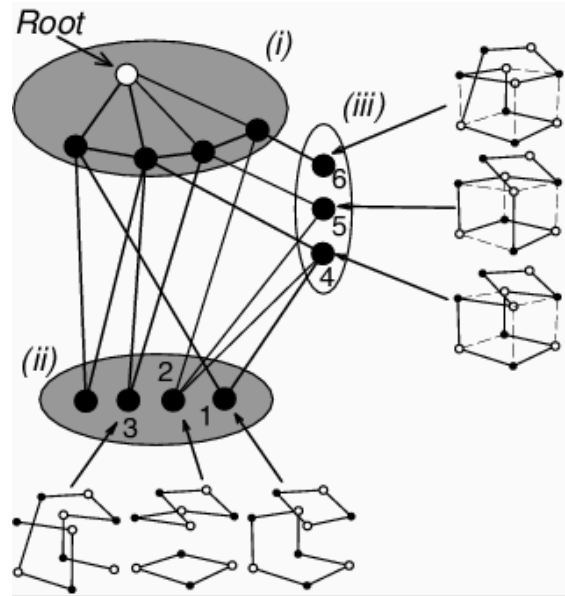}}}
\caption{The space of possible  configurations of links on the
cubic lattice. Different subspaces and example configurations
of links are shown.} \label{fig:graph}
\end{figure}

Let us  consider the procedures of  labeling random links,
branching and patching of algorithm \cite{Conf-des}. The random
labeling of links and branching of sub-chains may lead either
directly to the formation of the Hamiltonian cycle from subspace
(iii), or to the formation of some configuration from the subspace
(ii). The latter situation is much more probable due to the size
of the subspace (ii) is much larger than the size of (iii).
Suppose the algorithm  generated  some configuration from (ii).
Now the \emph{patching} procedure has to transform it to the
single cycle. Even if one supposes that configurations from (ii)
and (iii) are generated with equal probability, it appears that
the number of paths leading from (ii) to different Hamiltonian
cycles in (iii) is different. This can be easily seen from the
enumeration of all possible ways to label the $2\times 2\times 2$
lattice. The configurations 1 and 2 can be transformed to the
Hamiltonian cycles 4 and 5, but there is no way to obtain the
cycle 6 as a result of patching. Moreover, the number of paths to
cycles 4 and 5 is also slightly different. In general, the
probability to generate some Hamiltonian walk is proportional to
the number of possible configurations of sub-cycles which can be
transformed to this walk and to the number of ways to apply
patching procedures to these configurations of sub-cycles. And
this is the patching procedure that leads to the biased sampling
of Hamiltonian walk. Figure \ref{fig:graph} gives a simple
example.

Also it can be shown that the formation of the configuration with
dead ends (similar to the configuration 3 in Fig. \ref{fig:graph})
produces biased sampling of Hamiltonian walks too. The dead end
forms if some vertex of the lattice  which has only one incoming
link has no unsaturated neighbors.

The algorithm \cite{Conf-des} can be corrected by avoiding, on all
stages, placing a new link if it leads to either the closing of a
sub-cycle, or the formation of the dead ends.  If the formation of
the sub-cycles and the dead ends is forbidden, then paths starting
from the root configuration and ending in the subspace (iii) do
not pass  through the subspace (ii), and the patching is not
applied.

Undoubtedly, placing the links on the lattice in random order does
not produce any biases.  As for the branching of the sub-chains we
are not so sure.  However, in our simulations we did not see any
worrisome signs from this procedure.

\begin{widetext}
\section{Pseudocode}
\begin{tabbing}
Output:a\= Case 1: Hamiltonian cycle WE on the extended lattice
graph;\kill
Input: \> A lattice graph $LG($vertices $V$,edges $E)$.\\
Output:\> Case 1: Hamiltonian cycle WE on the extended lattice
graph; \\
\>Case 2: (If $LG$ is even):Hamiltonian cycle $WL$ on $LG$.\\
\\
Begin;\\
aaaa\=\kill
\>Color vertices of $LG$ alternatively white and black;\\
\>(if Case 1): Generate extended lattice graph $EG$;\\
\>$PerformRandomBipartiteMatching();$\\
End.\\
\\
Subroutines:\\
aaaa\=aaaa\=aaaa\=aaaa\=\kill
\>$PerformRandomBipartiteMatching():$\\
\>Begin;\\
\>\>While(number of unsaturated vertices $>0$)\\
\>\>\>Choose random unsaturated vertex $P$;\\
\>\>\>Choose random neighbor $Q$;\\
\>\>\>if ($Q$ unsaturated):\>\\
\>\>\>\>$TryLinkVertices(P,Q)$;\\
\>\>\>else if ($Q$ saturated):\\
\>\>\>\>Choose direction along sub-chain, $QS$;\\
\>\>\>\>Find end of sub-chain, $T$;\\
\>\>\>\>$TryGrowSubchain(T)$;\\
\>\>\>\>Remove link $QR$;\\
\>\>\>\>$TryLinkVertices(P,Q)$;\\
\>\>\>End if;\\
\>\>End while;\\
\>End.\\
\end{tabbing}
\begin{tabbing}
aaaa\=aaaa\=aaaa\=aaaa\=\kill
\>$TryLinkVertices(P,Q)$:\\
\>Begin;\\
\>\>Draw link  $PQ$;\\
\>\>Find dead ends and cycles; \\
\>\>if dead ends found, or (length of cycle $<$ length of complete
Hamiltonian walk):\\
\>\>\> Remove link $PQ$;\\
\>End.\\
\>$TryGrowSubchain(T)$;\\
\>Begin;\\
\>\>List unsaturated neighbors of $T$;\\
\>\>While List is not empty:\\
\>\>\>Choose random vertex $X$ from List;\\
\>\>\>$TryLinkVertices(X,T)$;\\
\>\>\>if link $XT$ is drawn:\\
\>\>\>\>End.\\
\>\>\>else: \\
\>\>\>\>Remove link $X$ from List; \\
\>\>End while;\\
\> End.\\
\end{tabbing}
\end{widetext}

\end{document}